\documentclass[aps,prd,preprint,superscriptaddress,nofootinbib,showpacs]{revtex4}
\usepackage{graphics}
\usepackage{epsfig}
\usepackage{latexsym}
\usepackage{colordvi}
\usepackage{amsmath}
\usepackage{amssymb}

\begin{document}
\preprint{\parbox[b]{1in}{ \hbox{\tt PNUTP-06/A20} \hbox{\tt
KIAS-P06040} }}

\title{Baryons in AdS/QCD }

\author{Deog Ki Hong}
\email[E-mail: ]{dkhong@pusan.ac.kr} \affiliation{Department of
Physics, Pusan National University,
             Busan 609-735, Korea}
             \affiliation{Asia Pacific Center for Theoretical Physics,  POSTECH, Pohang 709-784, Korea}

\author{Takeo Inami}
\email[E-mail: ]{inami@phys.chuo-u.ac.jp} \affiliation{Department
of Physics, Chuo University, Tokyo 112-8551, Japan}

\author{Ho-Ung Yee}
\email[E-mail: ]{ho-ung.yee@kias.re.kr} 
\affiliation{School of Physics, Korea Institute for Advanced
Study, Seoul 130-012, Korea} \vspace{0.1in}

\date{\today}

\begin{abstract}
We construct a holographic model for baryons in the
context of AdS/QCD and study the spin-$1\over 2$ nucleon spectra and its
 couplings to mesons, taking full account of the effects from the chiral
symmetry breaking. A pair of 5D spinors is introduced  to represent
both left and right chiralities. Our model contains two adjustable
parameters, the infrared cutoff and the Yukawa coupling of bulk
spinors to bulk scalars, corresponding to the order parameter of
chiral symmetry. Taking the lowest-lying nucleon mass as an input,
we calculate the mass spectrum of excited nucleons and the nucleon
couplings to pions. The excited nucleons show a parity-doubling
pattern with smaller pion-nucleon couplings.

\end{abstract}

\maketitle

\newpage
\section{Introduction}
Quantum chromodynamics (QCD) is now firmly believed to be the theory
of strong interactions. It describes the strongly interacting
particles at high energies or in deep-inelastic scattering extremely
well, where the weak coupling perturbation is reliable. QCD at low
energies is on the other hand strongly coupled and highly nonlinear.
So far it has precluded any analytic solutions. The test of QCD at
low energies, therefore,  has relied on either indirect evidences,
based on chiral perturbation theory, or on lattice calculations for
limited cases. Any reliable approximate methods to solve QCD deserve
a careful study, since we may gain physical insight from them into
the low energy dynamics of QCD.

Recent discovery of the correspondence  between a string theory in
anti-de Sitter (AdS) space and  the conformal field theory (CFT)
in its boundary has established a new paradigm for understanding strongly
coupled gauge theories
\cite{Maldacena:1997re,Gubser:1998bc,Witten:1998qj}. The
holographic dual of strongly coupled gauge theory corresponds to
the low energy limit or the supergravity limit of string theory in
the large 't~Hooft coupling limit, which therefore provides a
solvable scheme for QCD. In this scheme, called AdS/QCD, taking the
number of color very large, one approximates QCD to be conformally
invariant and thus its coupling and the anomalous dimensions of
composite operators are taken to remain constant for all scales
higher than the QCD scale, $\Lambda_{\text{QCD}}$, which is then
introduced by hand.
AdS/QCD will  break down at high energies, where QCD is
asymptotically free. However, it was shown to give reasonable
results for the meson sector of low energy QCD
\cite{Babington:2003vm,Kruczenski:2003uq,Sakai:2004cn,Erlich:2005qh,DaRold:2005zs,Evans:2005ip,Ghoroku:2005vt}.
This scheme has also been successfully applied for technicolor theories
\cite{Hong:2006si,Hirn:2006nt,Piai:2006hy,Csaki:2003zu,Burdman:2003ya}, which are otherwise
very hard to solve.

As baryons are essential ingredients in QCD, it is important to
extend the above models for mesons to include baryons. In this paper
we construct a model for low-lying spin-$1\over 2$ baryons in the
context of AdS/QCD and study their spectra  and couplings to mesons,
taking full  account of the effects from chiral symmetry breaking.
Some efforts have been already made to study the baryon spectrum of high
spin/mass resonances~\cite{deTeramond:2005su}. 
However, the effects of chiral symmetry breaking in QCD
were not exploited. Since such effects are important for low
spin/mass baryons, our analysis is a first step to resolve the
discrepancy. We also point out that the recent model of Sakai and
Sugimoto from string theory \cite{Sakai:2004cn} contains baryons as
solitonic objects, though its detailed study has not been done yet.
We discuss the relation between our approach and theirs in section
V.

\section{Holographic baryons and chiral symmetry}

The low-lying states of baryons form an even-parity octet or
decuplet under the flavor ${\rm SU}(3)$ symmetry  for three light
$u,d,s$ quarks. The excited states of baryons form singlets, octets or
decuplets but nearly degenerate in parity. One can in principle
calculate in QCD the spectra of ground states and radially excited
states of the baryon octet and decuplet from the correlation
functions of following composite operators,
\begin{equation}
B^i_j=\frac{1}{\sqrt{3}}
\left(q^iq^kq^l\epsilon_{jkl}-\frac{1}{3}\delta^i_jq^mq^kq^l\epsilon_{mkl}\right)
\quad
\text{and}\quad\Delta^{ijk}=\frac{1}{\sqrt{6}}\left(q^iq^jq^k+\,{\rm
cyclic\,\,\,perm.}\right), \label{baryon}
\end{equation} which have
the same quantum number of the baryon octet and decuplet,
respectively. (The orbitally excited states will contain positive
powers of derivatives.) In Eq.~(\ref{baryon}) the italic indices
denote the quark flavors $u,d,s$ and we have suppressed both Dirac
indices and color indices of quarks, keeping the color indices
completely anti-symmetric to form a color singlet.

According to the AdS/CFT correspondence, the spin-$\frac{1}{2}$
baryon operator at the boundary will correspond to  a bulk  Dirac
field \cite{Henningson:1998cd}, while the spin-$\frac{3}{2}$
baryon operator to a bulk  Rarita-Schwinger field
\cite{Volovich:1998tj}. Note that spinors in 5D bulk are always
Dirac spinors. In this paper we consider the spin-$\frac{1}{2}$
baryons only for simplicity.
It is straightforward to extend our analysis to spin-$\frac{3}{2}$ baryons.
The key idea in our model is to introduce a pair of bulk
spinors to describe the boundary Dirac spinor. This
predicts the parity-doublet pattern of excited baryons, explaining the observed
parity-doubling in excited baryons
from the holographic picture even though the ground state still breaks the chiral
symmetry.\footnote{
{See~\cite{Jaffe:2005sq} for a similar explanation on parity-doubling.}}

The ground state spin-$\frac{1}{2}$ baryons will be
massless if the chiral symmetry of QCD is unbroken.\footnote{For two
flavors the massless spin-$\frac{1}{2}$ baryons are fundamental
under the flavor symmetry $SU(2)_L\times SU(2)_R$ and saturate the
axial anomaly as originally shown by 't Hooft~\cite{'tHooft:1980xb}. However for three
flavors they do not saturate the anomalies of the microscopic
theory. One therefore suspects the chiral symmetry has to be
broken for three flavors.} As they are composed of chiral quarks,
their left-handed and right-handed  components separately
fall into $(8,1)$ and $(1,8)$ or $(3,\bar 3)$ and $(\bar 3,3)$ under
the chiral symmetry ${\rm SU}(3)_L \times {\rm SU}(3)_R$. Another
way of describing this is to introduce a pair of spin-$1\over 2$ chiral
baryon operators $\cal{O}_{\rm L}$ and $\cal{O}_{\rm R}$ which
transform as $(8,1)$ and $(1,8)$ or $(3,\bar 3)$ and $(\bar 3,3)$
under chiral symmetry.
 A natural mechanism for nucleon masses would  be then the mass
coupling between them by both explicit and spontaneous chiral symmetry breaking.

The above picture for the effects of chiral symmetry breaking on
the baryon spectra works well  for the up and down quarks
but not for the strange quark 
due to the large strange quark mass.
In this paper we therefore focus  on
the two-flavor case with 
vanishing current quark mass,
where the effect from the quark-antiquark condensate dominates.

To account for the chiral symmetry and its breaking in holographic
QCD, we adopt the model in~\cite{Erlich:2005qh,DaRold:2005zs}
and take a slice of the AdS metric as
\begin{equation}
ds^2=\frac{1}{z^2}\left(-dz^2+\eta^{\mu\nu}dx_{\mu}dx_{\nu}\right),
\quad \epsilon\le z\le z_m\,,
\end{equation}
where $z_m=1/\Lambda_{\rm QCD}$ is the IR boundary and
$z=\epsilon\ll1$  is  the UV boundary, and $\eta^{\mu\nu}={\rm diag}(+---)$ is the
four-dimensional Minkowski metric. We introduce a bulk scalar
field $X_i^j$, bi-fundamental under $SU(2)_L\times SU(2)_R$, which
corresponds to the composite operator $\bar q_{iL}q^j_R$ at the
boundary, described by the action
\begin{equation}
S_X=\int {\rm d}^5x\sqrt{g}\,{\rm
Tr}\left[\left|DX\right|^2-M_5^2\left|X\right|^2-
\frac{1}{2g_5^2}\left(F_L^2+F_R^2\right)\right]
\end{equation}
where $M_5$ is the mass of the bulk scalar field $X$ and the
covariant derivative
$D_{M}X=\partial_{M}X-i{A_{L}}_{M}\,X+iX\,{A_{R}}_{M}$. The bulk
mass is determined by the
relation~\cite{Gubser:1998bc,Witten:1998qj},
$\Delta_0\,(\Delta_0-4)=M_5^2$, where $\Delta_0$ is the scaling dimension
of the boundary operator $\bar q_{iL}q^j_R$. (For QCD
$\Delta_0=3$.) $F_L$ and $F_R$ are the field strength tensors of
the ${\rm SU}(2)_L\times {\rm SU}(2)_R$ bulk gauge fields $A_L$
and $A_R$, respectively. We take the five dimensional coupling $g_5=2\pi$
to match the two-point correlation functions of vector currents in
QCD at the ultra-violet (UV)
region~\cite{Erlich:2005qh,DaRold:2005zs}.

When evaluated for the classical solutions to the bulk
equations of motion, the vector and axial vector gauge
fields, defined as $V=(A_L+A_R)/\sqrt{2}$ and
$A=(A_L-A_R)/\sqrt{2}\,$, at the UV boundary couple to vector and
axial vector currents $J_{V}^{a\mu}=\bar q \gamma^{\mu}t^a\,q$ and
$J_{A}^{a\mu}=\bar q \gamma^{\mu}\gamma_5t^a\,q$ of quarks,
respectively.
(The ${\rm SU}(2)$ generators
$t^a=\frac{1}{2}\sigma^a$, where  $\sigma^a$'s are Pauli matrices.)
The classical solution of the bulk field $X$ is given as
\begin{equation}
X_0(z)=\frac{1}{2}M\,z+\frac{1}{2}\sigma\,z^{3},\label{cond}
\end{equation}
where the constants $M$ and $\sigma=\left<{\bar q}_Lq_R\right>$
correspond to the bare mass and the chiral condensate of QCD,
respectively. In the holographic dual of QCD  the axial
gauge fields therefore get a 5D mass
\begin{equation}
m_A^2=g_5^2\,\frac{|X_0|^2}{z^2}\,,
\end{equation}
where we take $M\to0$ in the chiral limit.
The spectra and interactions from this theory have
successfully reproduced the observed low energy meson
physics~\cite{Babington:2003vm,Kruczenski:2003uq,Sakai:2004cn,Erlich:2005qh,DaRold:2005zs,Evans:2005ip,Ghoroku:2005vt}.

Since the spin-$1\over 2$ chiral baryon operators $\cal{O}_{\rm L}$ and $\cal{O}_{\rm R}$
have different representations under ${\rm SU}(2)_L \times {\rm SU}(2)_R$,
it is not possible to assign a single 5D spinor in the bulk for both of them.
We must instead introduce a pair of 5D spinors $N_1$ and $N_2$, corresponding to
$\cal{O}_{\rm L}$ and $\cal{O}_{\rm R}$, respectively, which transform as
$(2,1)$ and $(1,2)$ under $SU(2)_L\times SU(2)_R$.

The bulk action for the Dirac
spinor $N_1$ (and similarly for $N_2$) is then given as
\begin{equation}
S_{N_1}=\int {\rm d}^5x\sqrt{g}\, \left\{ \frac{i}{2}\bar
N_1e_A^M\Gamma^A\nabla_MN_1-\frac{i}{2}\left(\nabla_M^{\dagger}\bar
N_1\right)e_A^M\Gamma^AN_1 -m_5\bar N_1N_1 \right\} \label{action}
\end{equation}
where  $m_5$ is the mass of the bulk spinor, $e^A_M$ is the
vielbein, satisfying $g_{MN}=e_M^Ae_N^B\,\eta_{AB}$, and the Dirac
matrices $\Gamma^A$'s ($A=0,1,2,3,5$) satisfy
$\left\{\Gamma^A,\Gamma^B\right\}=2\eta^{AB}$. Fixing the gauge for
the Lorentz transformation, we take the vielbeins
\begin{equation}
e_M^A=\frac{1}{z}\eta_M^A\,.
\end{equation}
In terms of the spin connection $\omega_M^{AB}$ and the gauge fields
we have introduced
the Lorentz and gauge covariant derivative
\begin{equation}
\nabla_M=\partial_M+\frac{i}{4}\omega_M^{AB}\Gamma_{AB}-i (A_L^a)_M t^a,
\end{equation}
where the Lorentz generator
$\Gamma^{AB}=\frac{1}{2i}[\Gamma^A,\Gamma^B]$ for spinor fields.
The non-vanishing components of the spin connection are
\begin{equation}
\omega_M^{5A}=-\omega_M^{A5}=\frac{1}{z}\delta_M^A\,.
\end{equation}

 By comparing the correlators for a free theory, or by looking at
the UV asymptotic of wave-functions, one finds the mass of
the (d+1)-dimensional bulk Dirac spinor, corresponding to a
boundary operator of scaling dimension $\Delta$,
\begin{equation}
m_5^2=\left(\Delta-\frac{d}{2}\right)^2\,.
\end{equation}
The sign of $m_5$ will be related to the chirality of the
boundary operator~\cite{Henningson:1998cd,Contino:2004vy}.
Extremizing the action (\ref{action}) with respect to  $\bar N_1$,
we get
\begin{equation}
\left(ie_A^M\Gamma^A\nabla_M-m_5\right)N_1=0 \quad\text{and}\quad
\left[\delta \bar
N_1 e_A^5\Gamma^A\,N_1\right]_{\epsilon}^{z_m}=0\,.
\label{boundary}
\end{equation}
Since the classical solution to the bulk equation of motion will be
a source to the corresponding boundary operator in the AdS/CFT
correspondence, when evaluated at the UV brane, it is convenient to
decompose the bulk spinor as, with the 4D chirality projectors,
\begin{equation}
N_1=N_{1L}+N_{1R}\,,
\end{equation}
where $i\Gamma^5N_{1L}=N_{1L}$ and $i\Gamma^5N_{1R}=-N_{1R}$. (The four dimensional
$\gamma^5=i\Gamma^5$.)
Then,
the boundary terms in (\ref{boundary}) become
\begin{equation}
\left[z\delta \bar N_{1L}N_{1R}-z\delta\bar
N_{1R}N_{1L}\right]_{\epsilon}^{z_m}=0\,. \label{boundary1}
\end{equation}

We now Fourier-transform the bulk spinor as
\begin{equation}
f_{1L,R}(p,z)\,\psi_{1L,R}(p)=\int~{\rm d}^4 x\,N_{1L,R}(x,z)e^{ip\cdot
x}\,,
\end{equation}
where the 4D spinors satisfy with $\psi_{1L}=\gamma^5\psi_{1L}$ and
$\psi_{1R}=-\gamma^5\psi_{1R}$
\begin{equation}
\!\not \! p\,\psi_{1L,R}(p)=|p|\,\psi_{1R,L}(p)\,
\end{equation}
and $f_{1L,R}$ satisfy
\begin{eqnarray}
\left(\partial_z-\frac{2+m_5}{z}\right)f_{1L}&=&-|p|\,f_{1R}\label{fl}\\
\left(\partial_z-\frac{2-m_5}{z}\right)f_{1R}&=&|p|\,f_{1L}\,.
\label{fr}
\end{eqnarray}
Normalizable solutions with eigenvalues $|p|$ and
appropriate boundary conditions that we will specify shortly
give us 4D propagating (Dirac) spinors $\psi_1$ with mass $|p|$.
For $|p|=0$, we may obtain a chiral fermion.
Eliminating $f_{1R}$ in (\ref{fl}) with Eq.~(\ref{fr}), we get
\begin{equation}
\left(\partial_z^2-\frac{4}{z}\,\partial_z+\frac{6+m_5-m_5^2}{z^2}\right)f_{1L}=-p^2f_{1L}\,.
\end{equation}
Similarly,
\begin{equation}
\left(\partial_z^2-\frac{4}{z}\,\partial_z+\frac{6-m_5-m_5^2}{z^2}\right)f_{1R}=-p^2f_{1R}\,.
\end{equation}
Near $z=\epsilon$ the solution behaves as
\begin{equation}
f_{1L}\simeq c_1(1+2m_5)z^{2+m_5}+c_2|p|z^{3-m_5}\quad,\quad f_{1R}\simeq c_1|p|z^{3+m_5}-c_2(1-2m_5)z^{2-m_5}\,.
\label{behavior}
\end{equation}
Note that each component of $f_{1L}$ comes along with a corresponding component in $f_{1R}$.

In our case the scaling dimension of
the baryon composite operator is $\Delta=9/2$ and thus $m_5=\pm5/2$.
In 5D  the spinor mass  term is real but has a sign ambiguity. If we
take $m_5>0$, $z^{2+m_5}$ is always normalizable, but $z^{2-m_5}$ is
not normalizable. On the other hand, if we take
$m_5<0$, $z^{2-m_5}$ is normalizable and $z^{2+m_5}$ is not
normalizable.
In this paper we choose the sign of $m_5$  such that
the left-handed chiral zero mode from $N_1$,
before introducing chiral symmetry breaking, corresponds to the massless left-handed nucleons.
Setting $|p|=0$ in (\ref{behavior}) and requiring a normalizable left-handed mode,
we see that $m_5={5 \over 2}$ must be positive for $N_1$.
We then have to take $m_5=-{5\over 2}$ for $N_2$ to get
a right-handed zero mode from $N_2$.

Another way of arriving at the same conclusion is the following.
$N_1$ is a holographic description of the left-handed chiral
operator ${\cal O}_{ L}$. For $m_5>0$, the dominant component as
we approach $z=\epsilon \ll 1$ is $f_{1R}\sim z^{2-m_5}$ and
according to AdS/CFT dictionary, this acts as a source for ${\cal
O}_{ L}$. Note that chirality of
$f_1$ is kinematically selected at the boundary. Since $f_1$ and
${\cal O}_{ L}$ have the same representation under the chiral
symmetry, the coupling must be $\bar{f_1}{\cal O}_{ L}+{\rm
h.c.}$, which does not vanish only if $f_1=f_{1R}$. Had $m_5$ been
negative, we would have $f_{1L}$ instead of $f_{1R}$ near the
boundary, which would give wrong chirality to ${\cal O}_{
L}$.

To allow a left-handed zero mode for $N_1$, we impose the following boundary conditions,
consistent with Eq.~(\ref{boundary1}),
\begin{equation}
 \lim _{\epsilon\to 0}N_{1L}(\epsilon)=0 \,\,({\rm normalizability})\quad\text{and}\quad N_{1R}(z_m)=0.
\end{equation}
The  remaining boundary conditions for $N_{1L}(z_m)$ and
$N_{1R}(\epsilon)$ are then determined by the equations of motion in
(\ref{boundary}), which become
\begin{eqnarray}
\left(\partial_z-\frac{2+m_5}{z}\right)N_{1L}+i\!\!\not\!\partial
N_{1R}=0\quad\text{and}\quad
\left(\partial_z-\frac{2-m_5}{z}\right)N_{1R}-i\!\!\not\!\partial
N_{1L}=0\,,
\end{eqnarray}
where $\!\not\!\partial=\gamma^{\mu}\partial_{\mu}$ and
$\gamma^{\mu}$'s are the 4D gamma matrices. The boundary conditions for $N_2$
are similar except for
interchanging $L \leftrightarrow R$ and $m_5 \leftrightarrow -m_5$.

The normalizable zero mode ($p^2=0$) for a free spinor with
$m_5>1/2$ is $f_{1L}=c_1\,z^{2+m_5}$ and the nonzero modes
($p^2\ne0$) satisfy a Bessel equation and
$f_{1L,R}(p,z)=c_1^{\,\prime}z^{5\over
2}\,J_{m_5\mp\frac{1}{2}}(|p|z)$, where $J_n(x)$ is the Bessel
function of the first kind. The spectra of the excited
Kaluza-Klein tower are given as the zeros of the Bessel function,
$J_{m_5+{1\over 2}}(|p|z_m)=0$, as we impose $N_{1R}(z_m)=0$.
(The spectrum of non-interacting fermionic operators has been
considered in~\cite{deTeramond:2005su,Contino:2004vy}, and for
the string theory context in~\cite{Argurio:2006my,Kirsch:2006he}.)

Similarly, we introduce another bulk spinor $N_2$ of mass $-m_5=-{5\over 2}$ for
the right-handed components of the boundary baryon operator. If we
take the normalizable modes of $N_{2R}\equiv\frac{1}{2}(1-i\Gamma^5)N_2$ not
to vanish at the IR brane but $N_{2L}(z_m)=0$, then they will have
exactly same spectra as those of the normalizable $N_{1}$, except
the chirality of the zero mode.

The 4D parity exchanges ${\cal O}_{ L}\leftrightarrow {\cal O}_{ R}$.
Hence its natural holographic extension in 5D would be flipping $N_1$ into $N_2$
and vice versa, together with the usual 4D parity operation.
It is also consistent with the fact that 5D spinor mass term flips its sign under the parity.
The even-parity massive baryons in the chiral symmetry
limit will therefore
correspond to $(N_{1}+N_{2})/\sqrt{2}$, while the odd-parity
baryons to $(N_{1}-N_{2})/\sqrt{2}$\,.

Since the bulk scalar flips
the 4D chirality, it naturally introduces a
gauge-invariant Yukawa coupling between two bulk spinors,
\begin{equation}
{\cal L}_{\rm int}\ni -g\left({\bar N}_2XN_1+{\rm h.c.}\right)\,,
\label{int}
\end{equation}
which becomes in the chiral limit $M\to0$
\begin{equation}
-\frac{1}{2}(\,g\,\sigma )z^3{\bar N}_2N_1+{\rm h.c.}\label{mass}
\end{equation}
The zero modes of $N_1$ and $N_2$ become massive through the Yukawa interaction
(\ref{int}) when the chiral symmetry is broken or $\sigma\ne0$. Otherwise they remain massless
in the chiral symmetry limit, consistent with  the 't~Hooft anomaly matching
condition.
They will form a nucleon doublet under the
isospin ${\rm SU}(2)_I \subset {\rm SU}(2)_L \times {\rm SU}(2)_R$.
The Yukawa interaction also mixes two towers of massive modes from each of $N_1$ and $N_2$, and
breaks their degeneracy into the parity doublet pattern.

\section{Numerical Analysis}

We have constructed a holographic QCD model for spin-$1\over 2$
baryons. The model contains two parameters, $z_m$ (the IR cutoff scale)
and $g\sigma$. Note that $\sigma$ enters independently in the
meson sector. We identify the lowest mass eigenstate as the
nucleon isospin doublet $(p,n)$.  We take its  mass $0.94 {\rm
~GeV}$ as an input to fix $g$, given $z_m$ and $\sigma$.
We first use the parameters of previous
analysis of the meson sector~\cite{Erlich:2005qh,DaRold:2005zs}, $z_m=(0.33~{\rm
GeV})^{-1}$ and $\sigma={\sqrt{2}\xi\over g_5\,z_m^3}$ with $3.4\le\xi\le4.0$
to calculate the spectrum of excited resonances.

We also treat $z_m$ as a parameter of our model and try to fit $z_m$ and $g$ from
the first resonance, Roper N(1440) of mass $1.44{\rm ~GeV}$, as well as the nucleons.
Our best fit is for $z_m=(0.205~{\rm GeV})^{-1}$ and $g=14.4$.
In this case, baryons are assumed to have a different IR cutoff scale than mesons.

Another important possibility that we don't pursue in this paper is to relax $\Delta={9\over 2}$
for the dimension of spin-${1\over 2}$ baryon operators.
We would certainly expect non-vanishing anomalous dimension in strongly coupled QCD.
Because we are not aware of any reliable method to calculate it  in our scheme,
it would be interesting to take it as an additional fitting parameter.

As $N_1$ and $N_2$ are coupled with each other, the Kaluza-Klein (K-K) mode equations become
\begin{eqnarray}
&&\left(
\begin{array}{cc}
\partial_z -{\Delta \over z} & -{1\over 2}g\sigma z^2 \\
-{1\over 2}g\sigma z^2 & \partial_z -{4-\Delta \over z}
\end{array}
\right)
\left(
\begin{array}{c} f_{1L} \\ f_{2L}
\end{array}
\right)=-|p|
\left(
\begin{array}{c} f_{1R} \\ f_{2R}
\end{array}
\right)\,, \nonumber\\
&&\left(
\begin{array}{cc}
\partial_z -{4-\Delta \over z} & {1\over 2}g\sigma z^2 \\
{1\over 2}g\sigma z^2 & \partial_z -{\Delta \over z}
\end{array}
\right)
\left(
\begin{array}{c} f_{1R} \\ f_{2R}
\end{array}
\right)=|p|
\left(
\begin{array}{c} f_{1L} \\ f_{2L}
\end{array}
\right)\,,
\label{kk_mode}
\end{eqnarray}
with the boundary conditions $f_{1R}(z_m)=f_{2L}(z_m)=0$.
Eigenfunctions should be normalizable with respect to the measure $dz \over z^4$ over $z\in [0,z_m]$.

TABLE I is the result of our numeric analysis using the shooting method.
If we adopt the value of $z_m=(0.33{\rm ~GeV})^{-1}$ from the meson sector, we do not seem
to reproduce observed spin-$1\over 2$ baryon spectrum well.
This may indicate baryons have a different IR cutoff than mesons, or
we should take  account of possible anomalous dimension to $\Delta={9\over 2}$.
\begin{table}[htb]
\begin{tabular}{|c|c|c|c|c|c|c|c|c|}
\hline
$z_m ({\rm GeV}^{-1})$& $g$ & (p,n)({\rm GeV}) & N(1440) & N(1535) & 3rd &  4th& 5th & 6th \\
\hline\hline
$(0.33)^{-1*}$ & 8.67 & $0.94^*$ &2.14 &2.24   & 3.25&3.30 &4.35 & 4.36 \\
\hline
$(0.205)^{-1}$ & 14.4 & $0.94^*$ & $1.44^*$ & 1.50 & 2.08 & 2.12 & 2.72 & 2.75 \\
\hline
\end{tabular}
\caption{Numerical result for spin-$1\over 2$ baryon spectrum. $*$
indicates an input and we used $\sigma=\frac{\sqrt{2}\xi}{g_5\,z_m^3}$
with $3.4\le\xi\le 4$.} \label{mass}
\end{table}

In the case of tunable $z_m$, we find $z_m=(0.205{\rm ~GeV})^{-1}$
and $g=14.4$ is the best fit, nicely fitting $1.44{\rm ~GeV}$ for
N(1440) and predicting $1.50{\rm ~GeV}$ for N(1535) quite well.
Note that this is also consistent with the parity of N(1440)  ($P=+1$)
and N(1535) ($P=-1$) since our holographic picture predicts a parity-doublet
pattern. After N(1535), the next parity-doublet resonance
is $(2.08, 2.12){\rm ~GeV}$, which is quite off from the data.
We suspect our assuming zero anomalous dimension might be responsible for  the
discrepancy.

In higher resonances, the mass splitting in parity doublets
will decrease as the mixing due to the chiral symmetry breaking
 would be negligible, compared to the large original K-K masses.
It is also worthwhile to comment that the fitted values of $g$ are
in the region of validity of our model. The effective mass from
Table~\ref{mass} has the maximum strength ${1\over 2}g\sigma z_m^3$,
whose value is less than or comparable to $m_5$.

\section{ Nucleon-pion coupling}
The low energy dynamics of hadrons is well described by chiral Lagrangian,
where the baryon interaction with pions  is
dictated by the chiral symmetry. If we expand the
chiral Lagrangian in powers of pions,  the interaction for
the nucleon doublet $N=(p,n)^T$ becomes
\begin{equation}
{\cal L}_{\pi NN}=\frac{g_A}{F_{\pi}}\bar
N\gamma^{\mu}\gamma_5{t^a}N\cdot\partial_{\mu}\pi^a+\cdots\,,
\label{chiral}
\end{equation}
where $g_A$ and $F_{\pi}$ are the nucleon axial coupling and the
pion decay constant, respectively.  Using the equation of motion, the
leading term in the interaction Lagrangian~(\ref{chiral}) can be
rewritten as
\begin{equation}
-ig_{\pi NN}\,\bar N\gamma_5\sigma^a\,N\,\pi^a\,
\label{pinn}
\end{equation}
with $g_{\pi NN}=g_Am_N/F_{\pi}$, the Goldberger-Treiman relation.
The nucleon-pion coupling is  measured accurately, $g_{\pi NN}\simeq13$, from
the low energy pion-nucleon scattering.

In our  model of holographic
QCD we have introduced bulk  gauge fields $A_{L,R}$ together with
bulk scalars $X$ and bulk spinors $N_{1,2}$.
Among the solutions to the equations of motion of those fields
the normalizable modes will be mapped to the physical states in QCD
by the AdS/CFT correspondence.
To identify the 4D fields correctly we need to diagonalize the 4D kinetic terms after
decomposing the bulk fields with the eigen modes of the 5D equations of motion.

Expanding the 5D Lagrangian  for the bulk gauge fields up to quadratic terms, we
have
\begin{eqnarray}
{\cal L}_{G}
&\simeq&-\frac{1}{2g_5^2z}\left(\partial_{\mu}V_{\nu}-\partial_{\nu}V_{\mu}\right)^2
-\frac{1}{2g_5^2z}\left(\partial_{\mu}V_5\right)^2-\frac{1}{g_5^2z}\partial_zV_{\mu}\partial^{\mu}V_5
+(V_M\leftrightarrow A_M)\,.\label{kinetic}
\end{eqnarray}
We see that there are cross terms in the 4D kinetic Lagrangian which mix
between 4D components of (axial) vector mesons and their 5th components.
The 5th components of 5D bulk gauge fields provide the longitudinal components of the 4D massive vector mesons.
The axial vector bosons mix further with the phase of the bi-fundamental scalar $X$,
which can be seen easily, if we expand its Lagrangian, taking $X(x,z)=v(z)e^{iP}$,
\begin{equation}
{\cal L}_X={\rm Tr}\left(|DX|^2-M_5^2|X|^2\right)=\frac{v^2}{z^2}\,{\rm Tr}\,
\left(\partial_{\mu}P-\sqrt{2}A_{\mu}\right)^2-\frac{v^2}{z^2}
{\rm Tr}\left(\partial_zP-\sqrt{2}A_5\right)^2+\cdots\,.
\end{equation}

Now we introduce gauge-fixing terms, which will remove the cross terms,
\begin{equation}
{\cal L}_{G.F.}=-\frac{z^4}{2\xi_V}\left[\partial^{\mu}V_{\mu}-\xi_Vz\partial_z\left(\frac{V_5}{z}\right)
\right]^2
-\frac{z^4}{2\xi_A}\left[\partial^{\mu}A_{\mu}-\xi_Az\partial_z\left(\frac{A_5}{z}\right)
+2\sqrt{2}g_5^2\frac{\xi_A}{z^2}\,v^2P
\right]^2
\end{equation}
In the unitary gauge ($\xi_{V,A}\to\infty$) $V_5$ and the
linear combination $z\partial_z(A_5/z)-2\sqrt{2}v^2g_5^2P/z^2$ become infinitely massive and
 decouple from the theory.
However, we note that its orthogonal combination of $A_5$ and $P$ remain massless,  if
\begin{eqnarray}
z\partial_z\left(\frac{A_5}{z}\right)-2\sqrt{2}\,\frac{v^2g_5^2}{z^2}P=0\,.
\label{ortho}
\end{eqnarray}
The kinetic terms for the bulk gauge and scalar fields become in the unitary gauge
\begin{eqnarray}
{\cal L}_{K}&=&-\frac{1}{2g_5^2}V^a_{\mu}\left(-g^{\mu\nu}\partial^2+\partial^{\mu}\partial^{\nu}\right)V^a_{\nu}
-\frac{1}{2g_5^2}A^a_{\mu}\left[-g^{\mu\nu}\left(\partial^2-\frac{g_5^2v^2}{z^2}\right)
+\partial^{\mu}\partial^{\nu}\right]A^a_{\nu}\nonumber\\
& &+\frac{z^4}{2g^5}\partial_{\mu}A^a_5\partial^{\mu}A^a_5+\frac{1}{2}v^2z^2\partial_{\mu}P^a\partial^{\mu}P^a
-\frac{1}{2}v^2z^2\left(\partial_zP^a-\sqrt{2}A^a_5\right)^2\,.
\label{pi_kinetic}
\end{eqnarray}
With the condition~(\ref{ortho}) the kinetic terms of $A_5$ and $P$ can be rewritten as
\begin{equation}
{\cal L}_K(A_5,P)=\frac{z^4}{2g_5^2}\left(\partial_{\mu}A_5\right)^2+\frac{z^8}{8g_5^2v^2}
\left[\partial_z\left(\frac{\partial_{\mu}A_5}{z}\right)\right]^2-
\frac{v^2z^2}{8g_5^4}\left[\partial_z\left(\frac{z^3}{v^2}\partial_z\left(\frac{A_5}{z}\right)
\right)-4g_5^2A_5
\right]^2
\label{a5_kin}
\end{equation}
Since the massless combination of $A_5$ and $P$ should map to pions of QCD, the Nambu-Goldstone bosons
associated with the spontaneous chiral symmetry breaking, we take an Ansatz
\begin{equation}
A_5(x,z)=f_0(z)\pi(x),
\end{equation}
where $\pi(x)$ is identified as the pion field. Then, the last term in (\ref{a5_kin}) should vanish and
\begin{equation}
\partial_z\left[\frac{z^3}{v^2}\partial_z\left(\frac{f_0}{z}\right)\right]-4g_5^2f_0=0\,.
\end{equation}
In order to have a canonical kinetic term for pions we take the normalization of $f_0(z)$ 
\begin{equation}
1=\int_0^{z_m}{\rm d}z\left[\frac{1}{2g_5^2z}f_0^2+\frac{z^3}{8v^2g_5^4}\left(
\partial_z\left(\frac{f_0}{z}\right)\right)^2\right]
\end{equation}



Once the pion fields are correctly identified, it is straightforward to find their couplings
to nucleons and their excited states.
In holographic QCD pions couple to nucleons through
the (bulk) gauge coupling and the Yukawa coupling Eq.~(\ref{int}),
\begin{equation}
{\cal L}_{\rm int}({\cal N},A_5)=-i\frac{z}{\sqrt{2}}
\bar{\cal N}\gamma^5A_5\tau_3{\cal N}-
\frac{gz^3}{2\sqrt{2}v}\bar {\cal N}\gamma^5\partial_z\left(\frac{A_5}{z}\right)\tau_2{\cal N}\,,
\end{equation}
where we used the relation (\ref{ortho}) and the Pauli matrices $\tau_{2,3}$ act on
the doublet of 5D spinors, ${\cal N}=(N_1,N_2)^T$.
The 5D spinors are decomposed in terms of the K-K modes of (\ref{kk_mode}) as
\begin{equation}
{\cal N}_L=\sum_n
\begin{pmatrix}
f^{(n)}_{1L}(z)\psi^{(n)}_L(x) \\
f^{(n)}_{2L}(z)\psi^{(n)}_L(x)
\end{pmatrix}
\,,
\quad
{\cal N}_R=\sum_n
\begin{pmatrix}
f^{(n)}_{1R}(z)\psi^{(n)}_R(x) \\
f^{(n)}_{2R}(z)\psi^{(n)}_R(x)
\end{pmatrix}
\end{equation}
with the normalization for each modes
\begin{equation}
\int_0^{z_m}\frac{{\rm d}z}{z^4}\left(\left|f^{(n)}_{1L}\right|^2+\left|f^{(n)}_{2L}\right|^2\right)=1
=\int_0^{z_m}\frac{{\rm d}z}{z^4}\left(\left|f^{(n)}_{1R}\right|^2+\left|f^{(n)}_{2R}\right|^2\right)\,.
\end{equation}
We then find the pion-nucleon-nucleon coupling for the lowest-lying nucleons
\begin{equation}
g_{\pi NN}=\int_0^{z_m}\frac{{\rm d}z}{2\sqrt{2}z^4}
\left[ f_0\left(
{f^{(1)}_{1L}}^*f^{(1)}_{1R}-{f^{(1)}_{2L}}^*f^{(1)}_{2R}\right)
-\frac{gz^2}{2vg_5^2}\,\partial_z\left(\frac{f_0}{z}\right)
\left({f^{(1)}_{1L}}^*f^{(1)}_{2R}-{f^{(1)}_{2L}}^*f^{(1)}_{1R}\right)
\right]
\end{equation}
Similarly one finds the couplings
for the pion-nucleon-Roper coupling and other couplings with excited states.
Numerical results for the couplings are presented in Table~\ref{table2}.

For parity-even states, the eigenfunctions have the properties
$f^{(n)}_{1L}=f^{(n)}_{2R}$ and $f^{(n)}_{1R}=-f^{(n)}_{2L}$, while parity-odd states
satisfy $f^{(n)}_{1L}=-f^{(n)}_{2R}$ and $f^{(n)}_{1R}=f^{(n)}_{2L}$.
The lowest-lying ($n=1$)
nucleons, (p,~n), are parity even, and the excited resonances appear as parity-doublet
$(N^+,N^-)$
with  the parity-even states ($N^+$) slightly lighter than the parity-odd states ($N_-$), as
shown in Table~{\ref{mass}.
These are all verified numerically.

Similarly to the pion coupling to the lowest-lying nucleons,
Eq.~(\ref{pinn}), the coupling of the pion to the nucleon $N$ and
the first-excited resonance $N(1440)$ is given as $ig_{\pi
NN_{1440}}\bar{N}\sigma^a \gamma^5 N_{1440}\pi^a+{\rm h.c.}$,
whereas the coupling to the parity-odd excitation $N(1535)$ as
$ig_{\pi N N_{1535}} \bar{N}\sigma^a N_{1530}\pi^a+{\rm h.c.}$.
Our K-K reduction reproduces precisely the above interactions that
are consistent with parity.

As we vary $z_m$ away from the fit from the meson sector, the
value of the chiral condensate $\sigma$ in (\ref{cond}) must be
determined by other means. We adopt the fit in~\cite{DaRold:2005zs},
$\sigma=\frac{\sqrt{2}\xi}{g_5\,z_m^3}$ with
$3.4\le\xi\le 4$ for all $z_m$.
\begin{table}[htb]
\begin{tabular}{|c|c|c|c|c|}
\hline
$z_m ({\rm GeV}^{-1})$& $\xi$ & $g_{\pi NN}$ & $g_{\pi NN_{1440}}$ & $g_{\pi N N_{1535}}$ \\
 \hline\hline
$(0.205)^{-1}$ &$3.4$ & 7.34 & -0.025 & 2.04 \\
\hline
$(0.205)^{-1}$ &$3.7$ & 6.52 & 0.30  & 1.88 \\
\hline
$(0.205)^{-1}$ &$4.0$ & 5.84 & 0.56  & 1.74 \\
\hline
\end{tabular}
\caption{Numerical results for  pion-nucleon-nucleon couplings. }
\label{table2}
\end{table}
TABLE II is our prediction for these couplings when $z_m=(0.205~{\rm GeV})^{-1}$.
For $z_m=(0.33{\rm
~GeV})^{-1}$ we find $g_{\pi N N}=4.35, 3.78, 3.30$
for $\xi=3.4,3.7,4.0$, respectively. 

\section{Relation to Baryons as Skyrmions}

In AdS/QCD we have gauge fields living in the bulk AdS whose dynamics is dual to the meson sector of QCD,
including pions and higher resonances. As is well-known, the pion sector can have
a topological skyrmion solution,
which is interpreted as a baryon,
and its fermionic character comes from a topological WZW term upon quantization.

The QCD skyrmion appears as an ``instanton particle" in the bulk
AdS~\cite{Son:2003et,Sakai:2004cn}. Neglecting the time direction, ${\rm
AdS}_5$ has 4 spatial directions ($R^3$+$z$-direction) and we can
have a gauge field configuration along these 4 directions which
simply looks like an ordinary instanton configuration. It is a
static, particle-like, Minkowski object in ${\rm AdS}_5$ spacetime. Its
fermionic nature is introduced by a 5d Chern-Simons term in ${\rm AdS}_5$,
which is again dual to the WZW term in QCD. These terms should exist
to match chiral anomalies. If we identify Nambu-Goldstone pions as Wilson
lines of the broken axial gauge fields, the
topological instanton charge in ${\rm AdS}_5$ is identical to the
winding number of skyrmions in QCD~\cite{Son:2003et,Eto:2005cc}, confirming
the correspondence.

In QCD, we may describe baryons as either skyrmions of the chiral
Lagrangian or we may introduce separate  baryon fields independently
and couple them to mesons. The
latter seems more effective and convenient. It is easier to
identify and realize chiral symmetry property in the latter
approach.

We can take the same effective approach in the holographic dual of QCD.
We describe ``instanton particles" with a 5d Chern-Simons
 term in ${\rm AdS}_5$ by effective local spinor  fields, and couple them
to gauge fields representing the meson sector of QCD. Aspects of
chiral symmetry and its breaking are more manifest in this
effective theory model as we have studied in this work.

Another justification for this approach is the following. As the
axial gauge symmetry is broken in ${\rm AdS}_5$ spacetime, its
instantons representing holographic baryons tend to shrink to zero
size, eventually go beyond the validity of our effective theory.
As they become point-like small instantons, we may introduce an
effective baryon field to describe them. In a more detailed model
like $D4$-$D8$-$\overline{D8}$ system \cite{Sakai:2004cn}, the
size of the instantons is stabilized, and the effective field
description would be reliable in low energy limit. It also
explains the appearance of two 5D spinors in our model; as we have
${\rm SU}(2)_L \times {\rm SU}(2)_R$ gauge fields, $N_1$ describes
${\rm SU}(2)_L$ ``instanton particles" and $N_2$ corresponds to
${\rm SU}(2)_R$ ``instanton particles". Their charges arise from
quantization of collective coordinates.

\section{Conclusion and outlook}
We have constructed a holographic model of QCD to study the spectrum of nucleons and their
coupling to pions. In our approach it is necessary to introduce a pair of bulk spinors
to describe baryons in QCD. The pair of spinors naturally interact
with each other through a Yukawa coupling by a bulk scalar, corresponding to the order parameter
of chiral symmetry breaking.

Our model has two adjustable parameters, the infrared cutoff $z_m$
and the Yukawa coupling $g$, with which we fit the baryon
spectrum. Taking the lowest-lying nucleon mass as an input, we
have calculated the mass spectrum of excited nucleons and their
couplings to pions, which are in a reasonable agreement with data,
though it deviates from the data quite a lot for highly excited
states. We have also calculated the pion-nucleon-nucleon
couplings. The pion coupling with lowest-lying nucleons is about
50\% less than the observed value, $g_{\pi NN}\simeq 13$. However
we expect our results will improve if we take account of the
anomalous dimension of the baryon operator and if we relax our
${\rm AdS}_5$ geometry to have a soft IR wall.

We find that our model  explains naturally the parity-doubling of
excited baryons. The mass difference between parity-doublet
nucleons and their couplings to pions become smaller for excited
nucleons, suggesting that the parity-doubling or the effective
chiral restoration in excited baryons is due to the holographic
nature of QCD or large $N_C$ QCD.

In our analysis we have focused on the spin-$\frac{1}{2}$ baryons in the two-flavor case.
It is however straightforward to extend our work to spin-$\frac{3}{2}$ baryons and to
the three-flavor case.

\vfill \eject

\acknowledgments We acknowledge useful discussions with Nakwoo Kim,
Youngman Kim, Shin Nakamura, Mannque Rho, Sang-Jin Sin and M.A.Stephanov. Two of
the authors (D.K.H. and H.U.Y.) would like to thank the APCTP for
organizing and sponsoring the focus program on ``Search for Exotic
State of Dense Matter" in 2006, where this work was partly done.
 The work of D.~K.~H. was supported by grant  No.
R01-2006-000-10912-0 from the Basic Research Program of the Korea
Science \& Engineering Foundation and H.~U.~Y. is partly supported
by grant No. R01-2003-000-10391-0 from the Basic Research Program of
the Korea Science \& Engineering Foundation.
The work of T.I. is supported by the grants of JSPS (Basic research B,C)
and the Ministry of Education and Science (priority area).

\end{document}